\documentclass[runningheads,citeauthoryear]{apinv40}
\usepackage{epsfig,cite,graphics}
\usepackage{marvosym} 
\usepackage[utf8]{inputenc}

\begin{document}

\title{Mass accretion rate in the jet-driving  
symbiotic binary MWC 560}
\titlerunning{The symbiotic star MWC 560}
\author{V. D. Marchev\inst{1}, R. K. Zamanov\inst{1}}
\authorrunning{Marchev \& Zamanov}
\tocauthor{V. D. Marchev, R. K. Zamanov} 
\institute{ $^1$ Institute of Astronomy and NAO, Bulgarian Academy of Sciences, 
                 Tsarigradsko shose 72, BG-1784 Sofia, Bulgaria     
} 
\papertype{ submitted on 24 August 2023; Accepted on xx.xx.xxxx }	
\maketitle

\begin{abstract}
We analyze photometric observations of the symbiotic  star MWC~560 in B and V bands 
obtained during the period 1990-2023. We  
estimate the luminosity and the mass accretion rate of the hot component. 
We find that the luminosity varies in the range from 200~$L_\odot$ to 3000~$L_\odot$, 
corresponding to a mass accretion rate in the range
$1 \times 10^{-7} - 2 \times 10^{-6} \; M_\odot \; yr^{-1}$ 
(for a  0.9~$M_\odot$ white dwarf and distance 2217 pc). 
The optical flickering disappears at  
mass accretion rate of about $1\times 10^{-6}\; M_\odot \; yr^{-1}$,
which sets an upper limit for the short-term variability from accreting white dwarf.
\end{abstract}
\vskip 0.3cm

\section{Introduction}

The symbiotic stars are wide binaries with long orbital periods
(from 100 days to 100 years),  in
which material is transferred from a red giant 
to a white dwarf or a neutron star (Miko{\l}ajewska 2012).

The outflow-driving star MWC~560 (V694 Mon) was identified
as an emission line object by Merrill \&  Burwell (1943) 
in the Mount Wilson observatory spectroscopic surveys. 
The spectroscopic observations of MWC~560 in 1984 showed  that it is an
extraordinary symbiotic star with absorption extending out to
$-3000$~\kms \ at H$\beta$, and other Balmer lines (Bond
et al. 1984). During  the spectroscopic observations at Rozhen Observatory in January - March 1990 
the outflow velocities reached 6000 -- 7000~km~s$^{-1}$, 
the absorption was well separated  from the emission. 
Tomov et al. (1990a) proposed that 
the absorption is caused by jets along the line of sight. 
The outflow may be a highly-collimated baryon-loaded jet (Schmid et al.
2001) or a wind from the polar regions (Lucy, Knigge \& Sokoloski 2018).
MWC~560 is considered to be a non-relativistic analog of the quasars
because 
 (i) collimated outflow (jets),
 (ii) the optical emission lines (Balmer lines and FeII lines) are similar 
      to those of the low-redshift quasars (Zamanov \& Marziani 2002),
and 
(iii) the absorption lines are similar to the lines of 
      the broad absorption lines quasars (Lucy et al. 2018).
The orbital period of the binary is probably $P_{orb} = 1931 \pm
162$~d (Gromadzki et al. 2007), although
Munari et al. (2016) proposed that it can be as short as  
$P_{orb} \approx 330.8$~d.

Here, we analyze B and V photometric observations 
and estimate the $(B-V)_0$ colour, effective temperature, 
luminosity and mass accretion rate of the hot component.  

\section{Observations}

We use B and V band data available in the AAVSO database as well as data from Tomov et al. (1990b) and Zamanov et al. (2020). In Fig.1 we plot the AAVSO data for V band (upper panel, green), B band (lower panel, blue). The red plusses are the season average magnitude 
for the respective observational season (from October to March). 
We take data for seasons where both B and V measurements are available 
because both are needed for our further calculations. 
In Table 1 we present the photometric magnitudes 
used in this work. Column one is the Julian day of the observation 
or the season average when using the AAVSO data. 
Column two and three are the B and V band magnitudes.
The other columns are described at the end of Sect.~\ref{Ma}. 

 \begin{figure}    
   \vspace{7.5cm}     
   \includegraphics{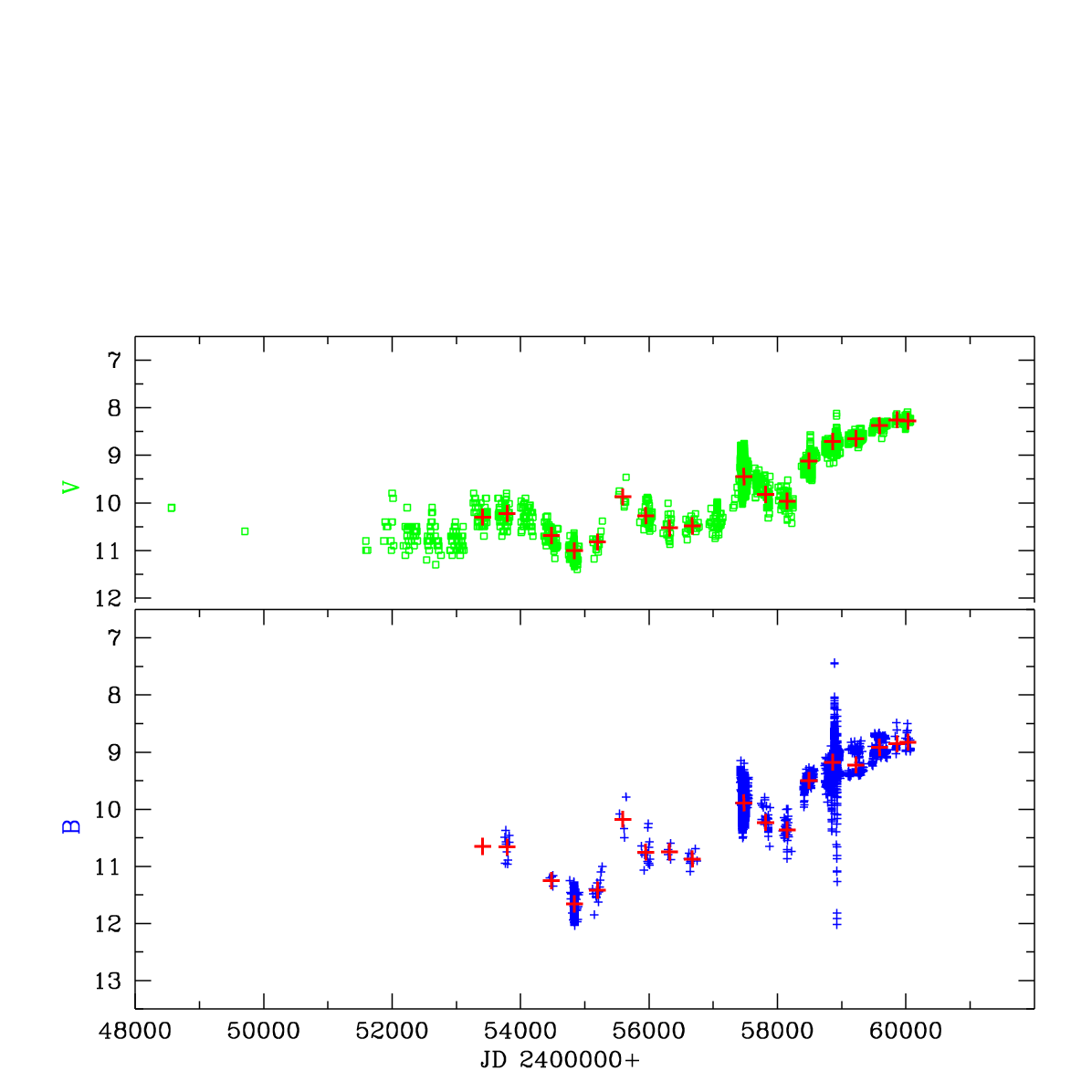}  
   \caption[]{The  AAVSO light curves of MWC 560 from 1987 to 2023 in B and V bands.
               The red plusses indicate the average value of the magnitude for each observational season.  } 
   \label{f.1} 
   \vspace{7.5cm}  
   \includegraphics{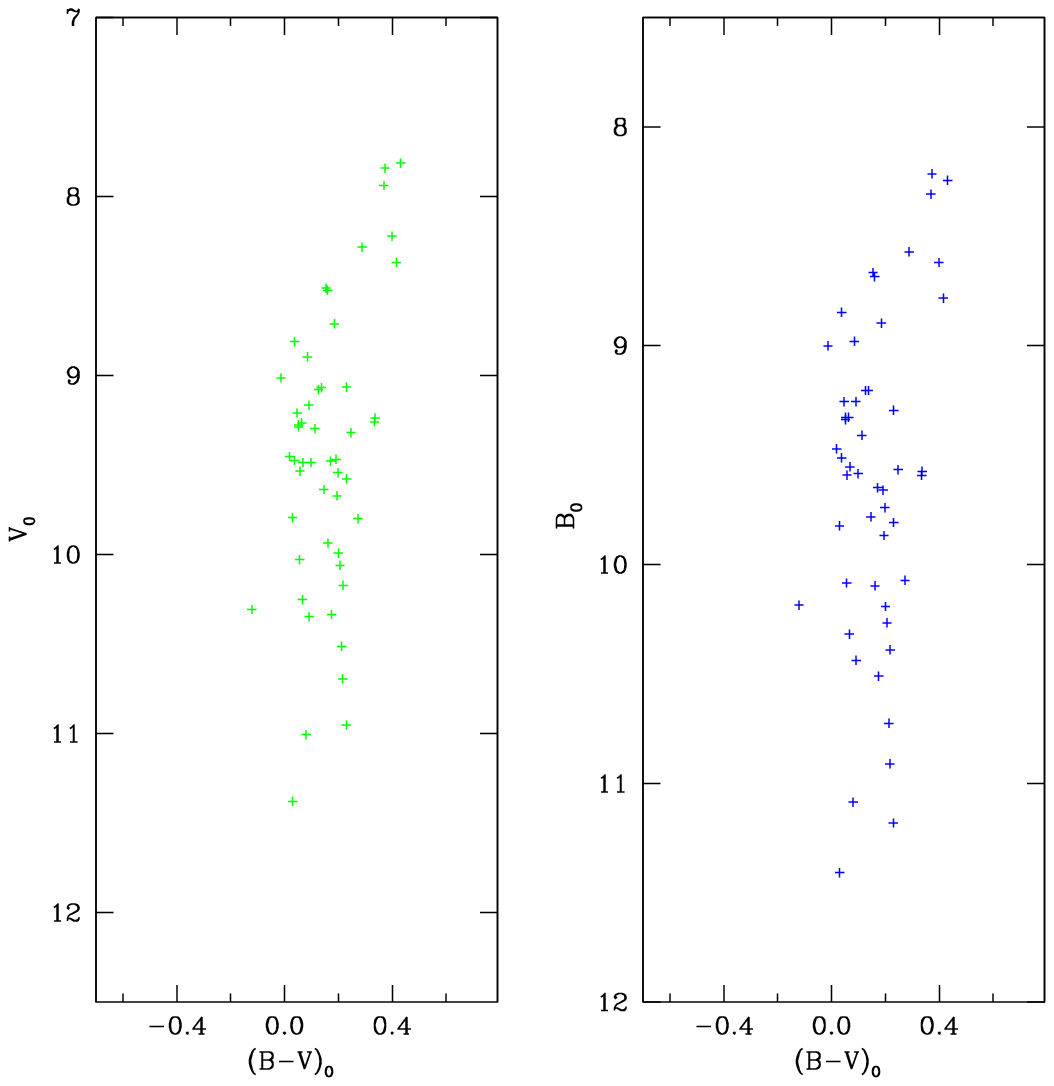}  
   \caption[]{Dereddened color-magnitude diagram for the hot component of MWC~560. 
              The left panel is $V_0$ versus $B-V_0$, 
	      the right panel is $B_0$ versus $B-V_0$. } 
   \label{f.cm}           
 \end{figure}	     

\begin{figure}[]
  \begin{center}
    \centering{\epsfig{file=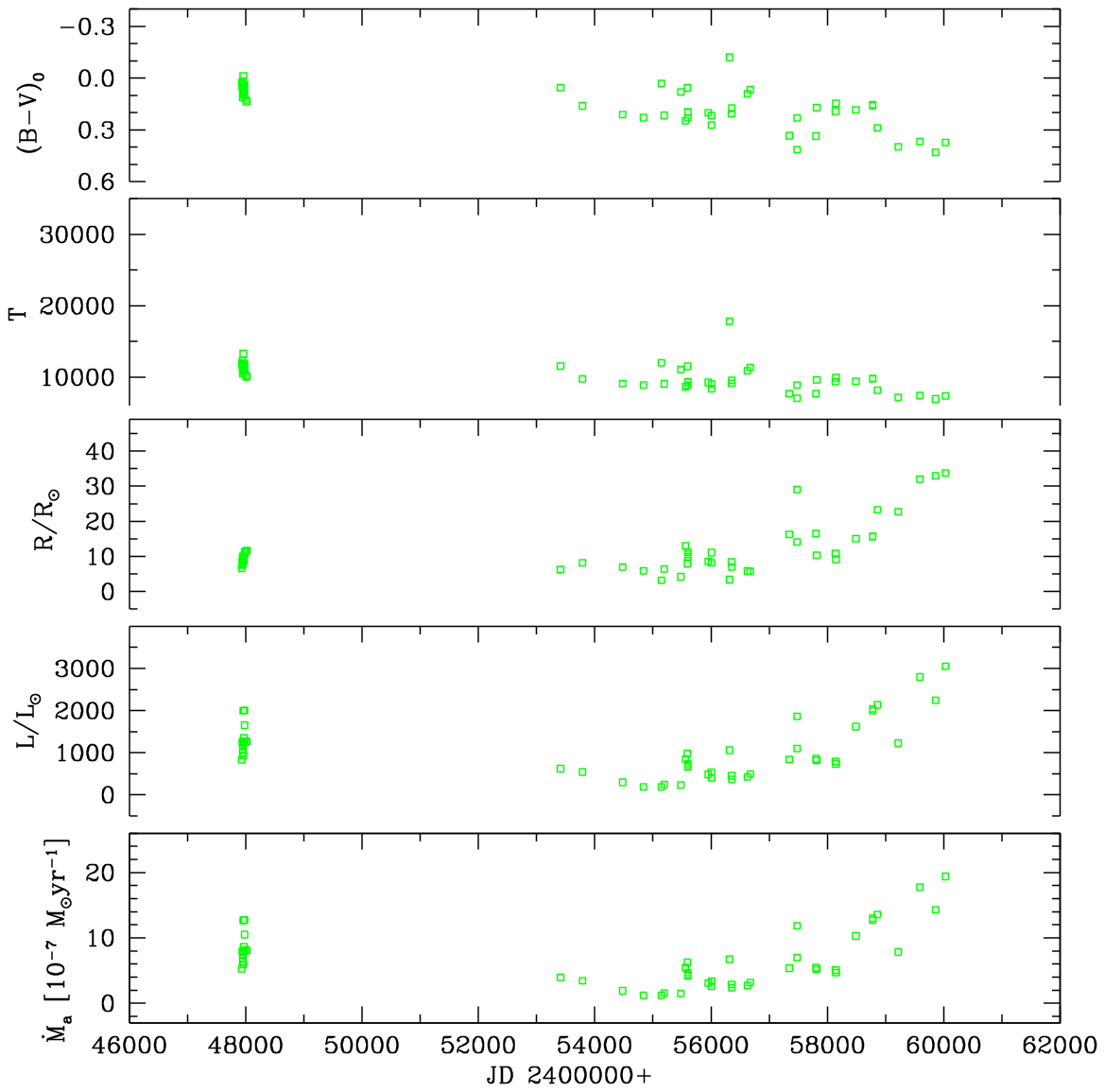, width=0.8\textwidth}}
    \caption[]{Evolution of the hot component of MWC~560.
     From top to bottom dereddened colour (B-V)$_0$, 
     effective temperature T[K], effective radius in solar units R/R$_\odot$, 
     luminosity in solar units L/L$_\odot$, and mass accretion rate M$_a$ [$10^{-7}\: M_\odot \: yr^{-1}$] 
     are plotted. }
    \label{f.d}
  \end{center}
\end{figure}

\section{Luminosity of the hot component}

To evaluate the brightness of the hot component, 
we subtract the red giant contribution and make the correction for the interstellar extinction. 
For the red giant we adopt apparent magnitudes $m_V =12.25$ and $m_B =13.94$ 
(see Sect.~3 in Zamanov et al. 2020). 
Schmid et al. (2001) estimated interstellar extinction  E(B-V)=0.15 mag
from the 2200~\AA\ feature. In the light of the NaD absorption and dust maps,  
the extinction is almost certainly in the range of $0.1 < E(B-V) < 0.2$ (Lucy et al. 2020).
Following the mean extinction law (Eq.1, Eq.3a, Eq.3b in Cardelli, Clayton, \& Mathis 1989), 
we estimate  
the extinction in B and V bands $A_B = 0.620$ and $A_V = 0.468$, respectively. 
The model of Bailer-Jones (2021) for the Gaia EDR3 data (Gaia Collaboration et al. 2018), 
gives a distance $d=2217$~pc  to MWC~560. 
To estimate the luminosity of the hot component, 
we applied the following  procedure:

1. From the apparent  magnitudes we subtract the contribution of the  red giant, 
using the calibrations given in Rodrigo et al. (2018):   
for Generic Bessell.B  filter -- 
effective wavelength  4371.07~\AA, zero magnitude star 
$6.13 \times 10^{-9}$ erg cm$^{-2}$ s$^{-1}$ \AA$^{-1}$,  
and for Generic Bessell.V filter  -- effective wavelength  5477.70~\AA,  zero magnitude star
$3.63 \times 10^{-9}$ erg cm$^{-2}$ s$^{-1}$ \AA$^{-1}$.
This gives the apparent  B and V magnitudes of the hot component.

2. We correct them for the interstellar extinction,  calculate $B_0$, $V_0$, and
the dereddened $(B-V)_0$ colour of the hot component (see also Fig.~2). 

3. Using  $(B-V)_0$ and  the calibration for black body (Table 18 in Strayzis 1992), 
we calculate the effective temperature of the hot component, $T_{eff}$.

4. Using distance $d=2217$~pc, and the dereddened magnitudes $B_0$ and $V_0$, 
we estimate the effective radius ($R_{eff}$) of the hot component. A simple program 
for the calculation of the radius is given in the Appendix. 

5. To derive the optical luminosity of the hot component we use 
the standard formula:
\begin{equation}
L = 4 \pi R_{\rm eff}^2 \; \sigma T_{eff}^4, 
\label{eqL}
\end{equation}
where $\sigma = 5.67 \times 10^{-5}$~erg cm$^{-2}$ s$^{-1}$ K$^{-4}$
is the Stefan-Boltzman constant. 
 
\section{Mass accretion rate onto the white dwarf}
\label{Ma}

The mass accretion rate onto the white dwarf is connected with the optical luminosity:    
\begin{equation}
L=\frac{1}{2} \; G \; \frac{M_{wd} \; \dot M_{a}}{R_{wd}}, 
\label{eqMa}
\end{equation}
where G is the gravitational constant, $\dot M_a$ is the mass accretion rate, 
$M_{wd}$ is the mass of the white dwarf, 
$R_{wd}$ is the radius of the white dwarf. 
The underlying assumption in this equation is that the 
disc luminosity is half of the total accretion luminosity.
The other half is emitted in  UV/X-rays by the boundary layer between 
the  accretion disc and the  white dwarf
(more details can be found in Chapter 6 of Frank et al. 2012).
We adopt for the white dwarf in MWC~560
mass $M_{wd}=0.9 M_\odot$ and radius $R_{wd}=6221$~km (Zamanov et al. 2011). Using these values and Eq.\ref{eqMa} we calculate the mass accretion rate.

The calculated values of the parameters for each observational data point are presented in Table 1. 
The column four of the table contains the colour of the hot component,
which is used to derive the effective temperature given in column five. 
Column six is the calculated effective radius in solar radii and column seven is 
the luminosity in solar luminosity units. 
Column seven gives  the calculated mass accretion rate
in units $10^{-7}\: M_\odot \; yr^{-1}$.
All the data from the last 5 columns of Table~\ref{table1} are presented in Figure~\ref{f.d}. In the figure 
the evolution of the parameters is plotted, from top to bottom, 
in the order they were obtained -- (B-V)$_0$, $T_{eff}$, $R_{eff}$, $L$ and $\dot M_{a}$. It is well visible that after JD~2458000 the $R_{eff}$, $L$ and $\dot M_{a}$ gradually increase. The optical luminosity goes from around 800 to 3000 L$_\odot$ and the mass accretion rate from 5 to 20 in units $10^{-7}\: M_\odot \: yr^{-1}$. The main source of uncertainty comes from the estimation of the red giant's magnitude when subtracting it from the total to get the hot component. The calculated typical uncertainties are $\Delta T_{eff}=\pm 500 \; K$, $\Delta R_{eff}=\pm 6-8\%$, and $\Delta L=\pm 4\%$. The uncertainty in the luminosity is smaller because the uncertainty for the radius and temperature contribute oppositely.  The main source of uncertainty in $\Delta \dot M_{a}$ is coming from the assumptions made in Eq.\ref{eqMa} and can be as big as a factor of 1.5.


\section{Discussion}

In the presented color magnitude
diagram (Fig.2), $(B-V)_0$ is in the range from 0.0 to 0.4, which corresponds to black body
temperature 7000 - 13000 K. There is only one exception 
(JD 2456319), which is probably an unusual state, or an observational error, and/or
incorrect subtraction of the red giant contribution.  

The luminosity of the hot component of MWC~560
is in the range 200 - 3000 L$_\odot$.   
The luminosities of the hot components in 18 symbiotic stars 
are estimated using the IUE spectra by Muerset et al. (1991).
The luminosity of the hot component of MWC~560 (see Table 1)
is similar to the luminosity observed in AG~Peg, Z~And, SY~Mus, AX~Per, V443~Her (see Table~5
in  Muerset et al. 1991). 

%
%

We find mass accretion rate onto the white dwarf of MWC~560 
in the range $1 \times 10^{-7} - 2 \times 10^{-6}$~M$_\odot$~yr$^{-1}$.  
It is considerably higher (two orders higher) than the estimated mass accretion rates 
in the cataclysmic variables, which are in the range 
$10^{-11} - 10^{-8}$~M$_\odot$~yr$^{-1}$ (see Table~5 in Pala et al. 2022).
Most probably this is 
because the mass donor (a red giant) in MWC~560
supplies more material than the red dwarfs in the cataclysmic variables. 

The accretion-induced variability  links 
young stellar objects, white dwarfs, and black holes (Scaringi et al. 2015).
Non-periodic variability on a time scale of $\sim 10$~minutes (flickering) 
is observed in many accreting white dwarfs (Bruch  2021). 
However, the  searches  for  flickering  from the accreting white dwarfs 
in  symbiotic stars and related objects (Sokoloski, Bildsten \& Ho \ 2001;  
Gromadzki et al. 2006; Stoyanov 2012)  have shown that
the optical flickering  is a  rarely detectable  phenomenon in symbiotic stars.
Among more than 300 known symbiotic stars (Akras et al. 2019),
only in 12 objects flickering activity is detected till now 
(a list can be found in Zamanov et al. 2021).

The flickering of MWC 560 is visible in all observations obtained 
between 1984 and May 2018 (e. g. Lucy et al. 2020 and references therein). 
The amplitude in U and B bands is in the range 0.1 -- 0.4 mag and the detected
quasi-periods are from 11 to 160 min (Tomov et al. 1996; Georgiev et al. 2022).
The short term variability (flickering) ceased in October 2018 (JD 2458400) 
and was not detected since then (Marchev et al. 2022, 2023).
Our results 
indicate that the flickering disappears when the mass accretion rate is 
$\ge 1.10^{-6}$~M$_\odot$~yr$^{-1}$. 
We propose two possibilities, when the mass accretion rate exceeds 
a certain value   
(1) the accretion disc becomes stable, and no fluctuations in the disc can be generated, 
or    
(2) around the accretion disc forms a "cocoon" (optically thick envelope),
    which reprocesses the radiation and smears the fluctuations. 
 
Many properties of the flickering can be explained with the fluctuating accretion disc model, 
where variations in the mass-transfer rate through the disc propagate toward the accreting compact object
(Lyubarskii 1997; Kotov et al. 2001, Scaringi et al. 2012).  
If the first possibility is real, our results set an upper limit 
($\approx 10^{-6}$~M$_\odot$~yr$^{-1}$) 
above which  
fluctuations can not be generated in accretion disc around white dwarf.  

%

\section{Conclusions}
We analyze photometric observations of the jet-driving symbiotic  star MWC~560 in B and V bands, 
obtained during the period 1990-2023. 
We estimate the luminosity and the mass accretion rate of the hot component. 
This is done by finding the $(B-V)_0$ colour of the hot component 
in the range 0.0 to 0.4, which corresponds to effective temperature 7000 - 13000~K.
We find that the luminosity varies in the range from 200~$L_\odot$ to 3000~$L_\odot$.
Adopting a 0.9~$M_\odot$ white dwarf and distance 2217~pc, we 
derive  mass accretion rate in the range $1.10^{-7} - 2.10^{-6}\; M_\odot\; yr^{-1}$. 

The disappearance of the optical flickering 
corresponds to a luminosity of the hot component 1600~$L_\odot$. 
Our results indicate an upper limit of the mass accretion rate 
(of about $1.10^{-6}\; M_\odot \; yr^{-1}$) above which 
no flickering can be observed from accretion disc around white dwarf.

\section*{Acknowledgments}
It is a pleasure to thank the anonymous referee for very
thoughtful and stimulating comments, which led to significant improvements of the paper.
We acknowledge with thanks the variable star observations from the AAVSO International Database contributed by observers worldwide and used in this research.


\begin{table}[]
  \begin{center}
  \caption{The estimated parameters of the hot component of MWC~560. In the columns are given 
  Julian Day ($^a$ Tomov et al. 1990, $^b$ AAVSO data, $^c$ Zamanov et al. 2021), 
  B and V band magnitudes of MWC~560, estimated $(B-V)_0$ of the hot component, 
  effective temperature, effective radius, optical luminosity, and the mass accretion rate.  }
  \begin{tabular}{crrr | rr rrrrr | c | ccc}
    JD        & B [mag]     & V [mag] & $\;$ & $\;$ & & B-V$_0$  & $T_{eff}$ & $R_{eff}  $ & & L/L$_\odot$ & $\dot M_{a}$  & & \\
2400000+      &             &         & $\;$ & $\;$ & &	         & [K]	     & $[R_\odot]$ & &             & [$10^{-7}\: M_\odot \: yr^{-1}]$ & & \\
47930$\; ^a$  &     9.823   &  9.793  & $\;$ & $\;$ & & 0.0294   &  12018   &   6.7       & &  829  & 5.27   &   & \\  
47946$\; ^a$  &     9.338   &  9.286  & $\;$ & $\;$ & & 0.0517   &  11605   &   8.8	  & &  1250 & 7.94   &   & \\
47947$\; ^a$  &     9.554   &  9.486  & $\;$ & $\;$ & & 0.0678   &  11313   &   8.2	  & &  1000 & 6.35   &   & \\
47948$\; ^a$  &     9.512   &  9.475  & $\;$ & $\;$ & & 0.0378   &  11858   &   7.8	  & &  1088 & 6.91   &   & \\
47950$\; ^a$  &     9.471   &  9.452  & $\;$ & $\;$ & & 0.0190   &  12330   &   7.5	  & &  1179 & 7.49   &   & \\
47953$\; ^a$  &     9.410   &  9.297  & $\;$ & $\;$ & & 0.1126   &  10498   &   9.9	  & &  1073 & 6.82   &   & \\
47954$\; ^a$  &     9.256   &  9.166  & $\;$ & $\;$ & & 0.0902   &  10905   &   10.0	  & &  1273 & 8.09   &   & \\
47955$\; ^a$  &     9.328   &  9.264  & $\;$ & $\;$ & & 0.0635   &  11391   &   9.1	  & &  1239 & 7.87   &   & \\
47960$\; ^a$  &     9.001   &  9.014  & $\;$ & $\;$ & & -0.0133  &  13299   &   8.4	  & &  1994 & 12.67  &   & \\
47964$\; ^a$  &     9.585   &  9.486  & $\;$ & $\;$ & & 0.0988   &  10749   &   8.8	  & &  930  & 5.91   &   & \\
47969$\; ^a$  &     9.256   &  9.209  & $\;$ & $\;$ & & 0.0465   &  11700   &   9.0	  & &  1359 & 8.63   &   & \\
47974$\; ^a$  &     9.328   &  9.275  & $\;$ & $\;$ & & 0.0525   &  11591   &   8.8	  & &  1260 & 8.00   &   & \\
47982$\; ^a$  &     8.980   &  8.896  & $\;$ & $\;$ & & 0.0844   &  11011   &   11.2	  & &  1655 & 10.51  &   & \\
47983$\; ^a$  &     8.848   &  8.810  & $\;$ & $\;$ & & 0.0376   &  11862   &   10.6	  & &  2007 & 12.75  &   & \\
48008$\; ^a$  &     9.205   &  9.079  & $\;$ & $\;$ & & 0.1260   &  10255   &   11.3	  & &  1275 & 8.10   &   & \\
48016$\; ^a$  &     9.205   &  9.068  & $\;$ & $\;$ & & 0.1368   &  10058   &   11.7	  & &  1260 & 8.00   &   & \\
53411$\; ^b$  &    10.084   & 10.028  & $\;$ & $\;$ & & 0.0561   &  11525   &   6.3	  & &  624  & 3.96   &   & \\
53789$\; ^b$  &    10.098   &  9.937  & $\;$ & $\;$ & & 0.1611   &   9736   &    8.2	  & &  543  & 3.45   &   & \\
54480$\; ^b$  &    10.726   & 10.513  & $\;$ & $\;$ & & 0.2124   &   9095   &    7.0	  & &  297  & 1.88   &   & \\
54839$\; ^b$  &    11.181   & 10.952  & $\;$ & $\;$ & & 0.2297   &   8879   &    5.9	  & &  194  & 1.23   &   & \\
55149$\; ^c$  &    11.408   & 11.378  & $\;$ & $\;$ & & 0.0303   &  11995   &    3.2	  & &  192  & 1.22   &   & \\
55198$\; ^b$  &    10.911   & 10.694  & $\;$ & $\;$ & & 0.2162   &   9048   &    6.4	  & &  250  & 1.59   &   & \\
55480$\; ^c$  &    11.086   & 11.006  & $\;$ & $\;$ & & 0.0801   &  11089   &    4.2	  & &  239  & 1.52   &   & \\
55559$\; ^c$  &     9.566   &  9.319  & $\;$ & $\;$ & & 0.2469   &   8664   &    13.0	  & &  854  & 5.43   &   & \\
55593$\; ^b$  &     9.591   &  9.534  & $\;$ & $\;$ & & 0.0573   &  11504   &    7.9	  & &  981  & 6.23   &   & \\
55602$\; ^c$  &     9.808   &  9.577  & $\;$ & $\;$ & & 0.2309   &   8864   &    11.1	  & &  686  & 4.36   &   & \\
55603$\; ^c$  &     9.867   &  9.672  & $\;$ & $\;$ & & 0.1948   &   9315   &    9.9	  & &  659  & 4.19   &   & \\
55604$\; ^c$  &     9.740   &  9.543  & $\;$ & $\;$ & & 0.1977   &   9279   &    10.5	  & &  740  & 4.70   &   & \\
55949$\; ^b$  &    10.192   &  9.992  & $\;$ & $\;$ & & 0.2002   &   9248   &    8.6	  & &  488  & 3.10   &   & \\
56007$\; ^c$  &    10.072   &  9.800  & $\;$ & $\;$ & & 0.2726   &   8342   &    11.1	  & &  535  & 3.40   &   & \\
56009$\; ^c$  &    10.390   & 10.172  & $\;$ & $\;$ & & 0.2179   &   9026   &    8.2	  & &  403  & 2.56   &   & \\
56319$\; ^b$  &    10.186   & 10.307  & $\;$ & $\;$ & &-0.1211   & 17839    &    3.4	  & &  1062 & 6.75   &   & \\
56354$\; ^c$  &    10.266   & 10.061  & $\;$ & $\;$ & & 0.2058   &   9178   &    8.4	  & &  454  & 2.88   &   & \\
56356$\; ^c$  &    10.510   & 10.335  & $\;$ & $\;$ & & 0.1751   &   9561   &    7.0	  & &  369  & 2.34   &   & \\
56625$\; ^c$  &    10.438   & 10.347  & $\;$ & $\;$ & & 0.0914   &  10884   &    5.8	  & &  428  & 2.72   &   & \\
56675$\; ^b$  &    10.318   & 10.252  & $\;$ & $\;$ & & 0.0658   &  11349   &    5.8	  & &  496  & 3.15   &   & \\
57344$\; ^c$  &     9.592   &  9.258  & $\;$ & $\;$ & & 0.3342   &   7715   &    16.3	  & &  843  & 5.36   &   & \\
57478$\; ^b$  &     9.295   &  9.065  & $\;$ & $\;$ & & 0.2303   &   8871   &    14.1	  & &  1101 & 6.99   &   & \\
57480$\; ^c$  &     8.783   &  8.368  & $\;$ & $\;$ & & 0.4153   &   7039   &    29.1	  & &  1863 & 11.83  &   & \\
57806$\; ^c$  &     9.574   &  9.238  & $\;$ & $\;$ & & 0.3367   &   7694   &    16.5	  & &  858  & 5.45   &   & \\
57816$\; ^b$  &     9.649   &  9.478  & $\;$ & $\;$ & & 0.1712   &   9610   &    10.3	  & &  816  & 5.19   &   & \\
58142$\; ^c$  &     9.660   &  9.468  & $\;$ & $\;$ & & 0.1918   &   9352   &    10.8	  & &  799  & 5.07   &   & \\
58151$\; ^b$  &     9.782   &  9.637  & $\;$ & $\;$ & & 0.1457   &   9929   &    9.2	  & &  734  & 4.66   &   & \\
58491$\; ^b$  &     8.897   &  8.712  & $\;$ & $\;$ & & 0.1849   &   9439   &    15.1	  & &  1620 & 10.29  &   & \\
58778$\; ^c$  &     8.666   &  8.512  & $\;$ & $\;$ & & 0.1540   &   9825   &    15.6	  & &  2042 & 12.97  &   & \\
58781$\; ^c$  &     8.684   &  8.524  & $\;$ & $\;$ & & 0.1604   &   9745   &    15.7	  & &  1999 & 12.70  &   & \\
58858$\; ^b$  &     8.571   &  8.283  & $\;$ & $\;$ & & 0.2882   &   8148   &    23.2	  & &  2136 & 13.57  &   & \\
59221$\; ^b$  &     8.620   &  8.221  & $\;$ & $\;$ & & 0.3986   &   7178   &    22.7	  & &  1230 & 7.81   &   & \\
59589$\; ^b$  &     8.307   &  7.938  & $\;$ & $\;$ & & 0.3687   &   7427   &    32.0	  & &  2797 & 17.77  &   & \\
59860$\; ^b$  &     8.244   &  7.813  & $\;$ & $\;$ & & 0.4312   &   6930   &    33.0	  & &  2246 & 14.27  &   & \\
60032$\; ^b$  &     8.214   &  7.841  & $\;$ & $\;$ & & 0.3731   &   7391   &    33.8	  & &  3054 & 19.40  &   & \\ 
  \end{tabular}
  \label{table1}
  \end{center}
\end{table} 

\clearpage 

\section*{Appendix}

Here we give a simple program used for the calculation of the radius 
of a black body located at a distance 2217~pc, 
with temperature 10000~K, dereddened B band magnitude 11.0. This program uses 
the IDL Astronomy Users Library (https://idlastro.gsfc.nasa.gov/):  \\ 
----------------------------

   wave=4378.12 
   
   mag3= 11.0
   
   T1= 10000.0
                                                  
   R= 10.0
   
   d1= 2217.0*3.0857e18 
   R1= R * 6.96e10
   
   flu1= R1/d1*R1/d1 * planck(wave,T1)  	    

   mag1=-2.5*alog10(flu1/6.293e-9)
   
   dmag= mag1 - mag3
   
   R3 = R*sqrt(10\^(dmag/2.5))
    
   print, R, mag1, mag3
   
   print, R3  \\
----------------------------- \\
In the above program are used 1~pc$ = 3.0857 \times 10^{18}$~cm, effective wavelength 
of the B band 4378.12~\AA, and 
flux of zero magnitude star in B band $6.293 \times 10^{-9}$~erg~cm$^2$~s$^{-1}$~\AA$^{-1}$. 
The program uses an initial value of R given by hand and it calculates the flux with the built in Plank function for the coresponding wavelength and temperature. Finally it derives the Radius R3 based on the difference in the magnitude. 

\end{document}